\documentclass[12pt]{article}
\usepackage{graphicx}
\usepackage{cite}

\newcommand{\beq}{\begin{equation}}
\newcommand{\eeq}{\end{equation}}

\begin{document}

\begin{titlepage}

\begin{center}


{\hbox to\hsize {\hfill UCB/PTH-02/47 }}
{\hbox to\hsize {\hfill  LBNL-51652 }}
\bigskip


{\Large \bf Extra Dimensions at the Weak Scale and Deviations from Newtonian Gravity}

\bigskip

{\bf Z. Chacko}$^{\bf a,b,d}$,
and {\bf E. Perazzi}$^{\bf c,e}$ \\

\bigskip

$^{\bf a}${\small \it Department of Physics, University of
California,
Berkeley, CA 94720, USA \\
\medskip
$^{\bf b}$ Theoretical Physics Group, Lawrence Berkeley National Laboratory, \\
Berkeley, CA 94720, USA \\
\medskip
$^{\bf c}$ Physics Department, Sloane Laboratory,\\
Yale University, New Haven, CT 06520, USA \\
\medskip
$^{\bf d}${\rm email}: zchacko@thsrv.lbl.gov \\
$^{\bf e}${\rm email}: elena.perazzi@yale.edu} \\

\bigskip

{\bf Abstract}

\end{center}
\noindent

We consider theories in which the Standard Model gauge fields propagate in extra
dimensions whose size is around the electroweak scale. The Standard Model quarks
and leptons may either be localized to a brane or propagate in the bulk. This class of
theories includes models of Scherk-Schwarz supersymmetry breaking and universal extra
dimensions. We consider the problem of stabilizing the volume of the extra dimensions. We
find that for a large class of stabilization mechanisms the field which corresponds to
fluctuations of the volume remains light even after stabilization, and has a mass in the
$10^{-3}$ eV range. In particular this is the case if stabilization does not involve
dynamics at scales larger than the cutoff of the higher dimensional Standard Model, and
if the effective theory below the compactification scale is four dimensional. The mass of
this field is protected against large radiative corrections by the general covariance of
the higher dimensional theory and by the weakness of its couplings, which are Planck
suppressed. Its couplings to matter mediate forces whose strength is
comparable to that of gravity and which can give rise to potentially observable
deviations from Newton's Law at sub-millimeter distances. Current experiments
investigating short distance gravity can probe extra dimensions too small to be
accessible to current collider experiments. In particular for a single extra dimension
stabilized by the Casimir energy of the Standard Model fields compactification radii as
small as 5 inverse TeV are accessible to current sub-millimeter gravity experiments.

\end{titlepage}

\renewcommand{\thepage}{\arabic{page}}
\setcounter{page}{1}

\section{Introduction}

In the last few years theories in which the Standard Model gauge fields propagate in
flat extra dimensions at or around the weak scale have attracted considerable
interest.  For early work on such models see, for example
{\cite{A2}}{\cite{AK}}{\cite{ABQ}}. In some of these theories the Standard Model
quarks and leptons are localized to a brane but in others they too propagate in the
bulk.  Theories of this type have been used to explain, among other things, the
fermion mass hierarchy {\cite{AS}}, the number of fermion generations {\cite{DP}},
the non-observation of proton decay {\cite{ADP}}, the observed abundance of dark
matter {\cite{ST}}, the small neutrino mass {\cite{MP}}, the weak mixing angle
{\cite{HN}} {\cite{CHP}} and the origin of supersymmetry breaking (see e.g.
{\cite{SS}}). The implications of some of these models for current and future
collider experiments have been investigated {\cite{ACD}} and used to put bounds on
the size of the extra dimensions. The current bound is about 300 GeV$^{-1}$.

In this paper we consider the problem of stabilizing the volume of the extra dimensions.
We find that in a large class of stabilization mechanisms the field which corresponds to
fluctuations in the volume of the extra dimensions remains light even after
stabilization, and can give sizable corrections to Newtonian gravity. This field is a
scalar in the four dimensional effective theory, which we will call the radion. The
inverse mass of the radion is typically in the sub-millimeter range. Its couplings
violate the equivalence principle, albeit weakly, and mediate a force between matter that
is comparable to that of gravity.

The existence of this light field is generic if volume stabilization does not involve
dynamics at scales higher than the cutoff of the higher dimensional Standard Model, which
we denote by $M_*$, and if the effective theory below the compactification scale is four
dimensional. The reason is not difficult to understand.  Consider a theory where the
gauge fields propagate in $D>4$ dimensions. Since the radion, which we denote by $V$,
emerges from the higher dimensional gravity sector its action in the four dimensional
effective theory has the form below.

\begin{equation}
\int d^4x \sqrt{-g} \left( 2M_4^2 R^{\left(4\right)} +
{\rm c} \frac{2M_4^2}{V^2} \partial_\mu V \partial^\mu V
- \Omega( V ) \right)
\end{equation}

Here c is a dimensionless number of order one that depends on the number of dimensions,
$M_4$ is the four dimensional Planck scale and $\Omega( V )$ is the potential for the
radion. If the dynamics which stabilizes the radion emerges from scales lower than or
of order the cutoff of the higher dimensional Standard Model then $<\Omega>$ is less
than or of order $M_*^D <V>$. We then expect the mass$^2$ of the radion to be less than
or
of order $\frac{M_*^D <V>}{M_4^2}$. The requirement that the Standard Model gauge
coupling be perturbative up to the scale $M_*$ implies that $M_*^{D-4} <V>$ is bounded
from above. We will denote the upper bound on this quantity by $n_D$. For five
dimensions $n_5 \approx 740$, while for 6 dimensions $n_6 \approx 4000$ {\cite{CLP}}.
For all $D>6$ $n_D < 30^{D-4}$.  The bound on $n_D$ implies that $M_*$ is not more than
two decades larger than the inverse radius of the extra dimensions if they are
toroidal. Since we are assuming that the extra dimensions are at or about the weak
scale this then implies a radion mass which is far below the electroweak scale. For
example for one extra dimension of inverse radius 500 GeV and the cutoff $M_*$ of order
1.5 TeV we find a radion mass of about .005 eV. A particle of this mass can mediate
forces between particles up to a distance of about 40 microns, which is within the
range of experiments searching for deviations from Newtonian gravity at short
distances. Note that such a small compactification radius is well beyond the range of
current collider experiments searching for universal extra dimensions.

In this analysis above we have implicitly assumed that the effective theory below the
scale $R^{-1}$ is four dimensional and that the Standard Model along with the radion is a
complete description of physics below this scale. If the hierarchy problem is solved at
low scales as in the case of large extra dimensions {\cite{ADD}} this analysis does not
necessarily apply. We will consider this possibility in more detail in a subsequent
section.

One may also worry that such a light radion mass would not be stable under radiative
corrections, which would raise its mass to the electroweak scale. However this is not the
case. The reason is that the radion mass is protected by higher dimensional general
covariance, because the radion is part of the higher dimensional gravity multiplet. Hence
it is not possible to write a radion mass term in the higher dimensional theory.
Therefore the radion mass in the four dimensional effective theory must emerge as a
finite effect, and loops which contribute to the radion mass will be cutoff at the
compactification scale $R^{-1}$, where $R$ is the radius of the extra dimensions. However
since the couplings of the radion in the effective theory are all Planck suppressed the
corrections to the radion mass$^2$ from quantum effects will also be suppressed by at
least two powers of $M_4$. Since the only other scale besides $R^{-1}$ available to make
up the dimensions of mass$^2$ is the cutoff $M_*$, and because radiative effects are loop
suppressed, we do not expect radiative corrections to overwhelm our classical result.
For example the correction to the radion mass from the Casimir energy of bulk fields is
of order $\frac{1}{16 \pi^2} \frac{R^{-4}} {M_4^2}$ which is much smaller than the
typical classical contribution $\frac{M_*^D <V>}{M_4^2}$.

From where does the leading
correction to Newtonian gravity arise? Consider
the coupling of the radion to the SU(3)
gauge fields in the four dimensional effective theory which takes the form
\begin{equation}
\int d^4x \sqrt{-g} \frac{V}{g_D^2} {\rm Tr} F_{\mu \nu} F^{\mu \nu}
\end{equation}
Clearly $\sqrt\frac{g_D^2}{<V>}$ is the effective four dimensional SU(3) gauge
coupling $g$ at the matching scale, which is approximately the radius of the
extra dimensions, $R = \frac{<V>^{\frac{1}{D-4}}}{2\pi}$.
Now the mass of a nucleon is (up to corrections which depend on the light
quark massses) proportional
to $\Lambda_{QCD}$, the scale where the SU(3) gauge coupling becomes strong.
Since $\Lambda_{QCD}$ depends on the radion VEV, the nucleon mass depends on the
radion VEV and therefore there is a coupling of the radion to nucleons.
The coupling of the radion to the nucleon in linear order
\begin{equation}
\frac{\partial{m_N}}{\partial<V>} \left(V - <V>\right) N \bar{N}
= m_N \frac{\partial{\;log\left(\Lambda_{QCD}\right)}}{\partial<V>}
\left(V - <V>\right) N \bar{N}
\end{equation}
Using the renormalization group dependence of $\Lambda_{QCD}$ on $<V>$ we
find that the
approximate strength of the radion nucleon coupling is given by
\begin{equation}
\frac{8\pi^2}{9 g^2} \frac{m_N}{<V>}\left(V-<V>\right) \bar{N} N
\end{equation}
where $g$ is the strong gauge coupling at the scale $R^{-1}$.
Normalizing the radion kinetic term canonically by defining
\begin{equation}
\Sigma = 2 \sqrt{c} \frac{M_4}{<V>} \left(V - <V>\right)
\end{equation}
we see that the coupling of the canonically normalized radion to the nucleon has the
form $y_N \Sigma \bar{N} N$. From the computation above $y_N \approx
\frac{3.7}{\sqrt{c}} \frac{m_N}{M_4}$. However we have ignored the effects of the
light quark masses. Including these has the effect of altering $y_N$ by a factor of
$(1 - \chi)$ where $\chi \approx 0.27 \pm .08$ \cite{KW}. Therefore our final
expression for $y_N$ is $(1 - \chi) \frac{3.7}{\sqrt{c}}\frac{m_N}{M_4}$. The radion
nucleon coupling mediates a potential between matter that has a strength
$\frac{y_N^2}{4 \pi r} {\rm exp} (- m_V r)$. Comparing this to the gravitational
potential $G_N \frac{m_N^2}{r}$, at short distances the Yukawa force is stronger by a
factor {\bf P} = $(1-\chi)^2 \left(\frac{110}{c}\right)$. For one extra dimension $c
= \frac{3}{2}$ and the ratio of the radion mediated force to the gravitational force
is about $39 \pm 8$. For two extra dimensions $c = 1$ and the ratio of the two forces
is about $59 \pm 12$. Violation of the equivalence principle is also proportional to
the masses of the light quarks, but is highly suppressed {\cite{KW}} and is only at
the level of a few parts in a thousand.

In this estimate we have neglected corrections to the four dimensional gauge coupling
from higher dimensional operators suppressed by powers of the cutoff $M_*$ and from
brane localized kinetic and interaction terms. While the effects from higher
dimensional operators do not significantly affect the result, the effects of brane
localized terms, though volume suppressed, cannot in general be neglected. A quick
estimate suggests that the ratio of the contributions to the four dimensional gauge
coupling from the bulk terms and the brane localized terms is approximately $(\pi R
M_*):1$ for one extra dimension and $(\pi R M_*)^2:1$ for two extra dimensions. For
two extra dimensions the effect of these terms is typically small. However for one
extra dimension and $R M_*$ of order 3 the corrections to {\bf P} are of order 10
percent. The value of {\bf P} can be used to distinguish theories with weak scale
extra dimensions from other theories that also predict deviations from Newtonian
gravity at sub-millimeter distances.  These include large extra dimensions
{\cite{ADD}}, the string dilaton{\cite{KW}} and other string moduli which only get
mass after supersymmetry breaking {\cite{FKZ}}{\cite{KZP}} {\cite{DG}} and axions
{\cite{BAS}}.  Weak scale extra dimensions stabilized by supersymmetry breaking and
their implications for sub-millimeter gravity experiments have been considered
earlier in {\cite{ADD'}},{\cite{ADPQ}}. For a more complete list of theories which
predict deviations from Newtonian gravity see {\cite{LCP}}.

In the following sections we investigate in detail the validity of our conclusions above
regarding the the radion mass, which were based on naive scaling arguments. We first
consider a specific model of radion stabilization based on the Goldberger-Wise mechanism
and show how our conclusions naturally emerge.  We then consider the dependence of our
results on the manner in which the hierarchy problem is solved. Finally we consider the
implications of our result for models in which the radion is stabilized using the
Casimir energy of the Standard Model in the bulk, and then we conclude.

\section{Volume Stabilization}

In this section we consider in detail a simple model with a single compact extra
dimension and show how our earlier conclusions naturally emerge once the volume of the
extra dimension is stabilized. The compact direction is orbifolded by $S^1/Z_2$ and
there are branes at the orbifold fixed points. We use the Goldberger-Wise {\cite{GW}} mechanism
which involves a bulk scalar field with sources on the two branes to stabilize the
setup. Our ansatz for the metric is of the form
\begin{equation}
ds^2 = g_{\mu \nu}(x) dx^{\mu} dx^{\nu} + r(x)^2 d{\phi}^2
\end{equation}
where $\phi$ is the coordinate along the compact direction and runs from $-\pi$ to $\pi$.
It can be shown (see e.g. {\cite{LS}}) that this ansatz reproduces the correct action for the
radion in the four dimensional effective theory. The relevant part of the
action for our system can be decomposed as
\begin{equation}
S = S_G + S_M
\end{equation}
where $S_G$ and $S_M$ are the gravitational and scalar field constituents
of the action respectively.
\begin{equation}
S_G = \int d^4 x d\phi \sqrt{-G} (2M^3 R^{\left(5\right)} - \Lambda_B) -
\delta(\phi)\sqrt{-g}\Lambda_1 - \delta(\phi - \pi)\sqrt{-g}\Lambda_2
\end{equation}
\begin{equation}
S_M = \int d^4 x d\phi \sqrt{-G}\frac{1}{2} (-\partial_M \psi \partial^M \psi - m^2
\psi^2) + \sqrt{-g}F(\psi) \delta(\phi) + \sqrt{-g}H(\psi) \delta(\phi -
\pi)
\end{equation}
Here $F(\psi)$ and $H(\psi)$ are in general arbitrary functions of
$\psi$. For simplicity we will take
\begin{eqnarray}
F(\psi) &=& \lambda_1 \psi \\
H(\psi) &=& \lambda_2 \psi
\end{eqnarray}

We obtain an approximate solution for the coupled matter gravity system using the
method of Goldberger and Wise. We first solve for the scalar field $\psi$ in the
background metric as a function of $r$. We then integrate out the extra dimension to
obtain a potential for $r(x)$ which can be used to stabilize the extra dimension.  The
equation of motion for the scalar field is given by
\begin{equation}
-r^{-1} \partial_5 \partial_5 \psi + m^2 r \psi = \lambda_1 \delta(\phi) + \lambda_2
\delta(\phi - \pi)
\end{equation}
The solution (for $\phi > 0$) has the form
\begin{equation}
\psi = A exp\left(mr\phi\right) + B exp\left(-mr\phi\right)
\end{equation}
where $A$ and $B$ are given by
\begin{eqnarray}
A &=& \frac{1}{2m}\frac{\lambda_1 exp\left(-mr\pi\right)
+ \lambda_2}{2sinh\left(mr\pi\right)} \\
B &=& \frac{1}{2m}\frac{\lambda_1
exp\left(mr\pi\right) + \lambda_2}{2sinh\left(mr\pi\right)}
\end{eqnarray}
Substituting
the solution for $\psi$ back into the action and performing the integration over the
extra dimension we find
\begin{equation}
S_M = \int d^4x \sqrt{-g} \frac{1}{4m}
\left( coth\left(mr\pi\right)\left(\lambda_1^2 + \lambda_2^2 \right) + 2 \lambda_1
\lambda_2 cosech\left(mr\pi \right) \right)
\end{equation}
Similarly we substitute the
expression for the metric into $S_G$ and perform the integration over the extra
dimension to obtain
\begin{equation}
S_G = \int d^4 x \sqrt{-g} \left(4 \pi M^3 rR^{\left(4\right)}
- 2 \pi r\Lambda_B - \Lambda_4 \right)
\end{equation}
with $\Lambda_4=\Lambda_1+\Lambda_2$.
We extremize the non-derivative terms
in $r$ in the action to find $<r>$. At the minimum
$\frac{\partial{V\left(r\right)}}{\partial{r}} = 0$ which leads to
\begin{equation}
\frac{\pi}{4}\left[\left(\lambda_1^2 + \lambda_2^2 \right) cosech^2 \left(mr\pi\right)
+ 2 \lambda_1 \lambda_2 cosech \left(mr\pi\right) coth \left(mr\pi\right)\right]
+ 2 \pi \Lambda_B = 0
\end{equation}
This admits a (local) stable minimum for positive $\Lambda_B$ at large $\pi mr \gg 1$
provided $\Lambda_B \ll \lambda_1^2, \lambda_2^2$ and the product
$\lambda_1 \lambda_2$ is negative.

To get the action for four dimensional gravity into the canonical form we
perform a Weyl rescaling $g_{\mu \nu} \rightarrow \frac{<r>}{r} g_{\mu \nu}$.
Then the gravity action becomes
\begin{equation}
\int d^4 x \sqrt{-g} 4 \pi M^3 rR^{\left(4\right)}  = \int d^4 x \sqrt{-g} 2 M_4^2 \left( R^{\left(4\right)} +
\frac{3}{2 V^2}\partial_{\mu}V \partial^{\mu}V \right)
\end{equation}
where the radion field $V = 2\pi r$ and the four dimensional Planck scale emerges as
$M_4^2 = 2\pi M_5^3 <r>$.
After putting the kinetic term for the radion into the canonical form it is
straightforward to solve for the mass of the radion.
\begin{equation}
m_V^2 \approx \frac{m \Lambda_B V^2}{M_4^2} \le \frac{\lambda_1 \lambda_2
V}{M_4^2}
\end{equation}
Now if the dynamics which stabilizes the radion does not involve dynamics at scales
above the cutoff of the higher dimensional gauge theory $M_*$ then $\lambda_1 \lambda_2
\le M_*^5$ which implies $m_V^2 \le \frac{M_*^5V}{M_4^2}$. This is in perfect
agreement with our earlier conclusion which was based on dimensional grounds.
However if $\lambda_1 \lambda_2 \gg M_*^5$ then the radion mass can be much heavier
than inverse millimeter. We consider the implications of this possibility in the next
section.

\section{The Hierarchy Problem}

Under what circumstances is the condition $\lambda_1 \lambda_2 \le M_*^5$ satisfied? This
is related to the manner in which the hierarchy problem is solved.  Consider the
situation in which the hierarchy problem is solved as in the Randall-Sundrum model
{\cite{RS}}. There is now a warped extra dimension in addition to the compact extra
dimension. Models of this type have been considered in {\cite{CN}}. We first consider the
situation where the scalar field is confined to the brane where the Standard Model fields
live. Then the cutoff on the infrared brane $M_*$ must be around the weak scale to solve
the hierarchy problem, and $\lambda_1 \lambda_2 \le M_*^5$. Assuming that the radius of
the warped extra dimension has been stabilized and has mass of order $M_*$
{\footnote{This will be the case if, for example, it is stabilized using the
Goldberger-Wise mechanism.}} we can integrate it out to obtain an effective theory which
has the same form as in the five dimensional model considered above, and therefore this
theory has a light radion. One may worry that mixing between the radion and the
excitation corresponding to the radius of the warped extra dimension may change this
result but we have checked that such mixing, though present, is too small to affect our
conclusions. If however the scalar field lives on the ultraviolet brane then $\lambda_1
\lambda_2$ can be as large as $M_4^5$ implying a radion much heavier than inverse
millimeter.  However we see from eqn. (18) that unless the mass $m$ of the scalar field
is around the weak scale the size of the extra dimension will tend to be at the Planck
scale. This suggests that there is a naturalness problem which needs to be addressed in
this framework if the radion is to be much heavier than inverse millimeter.

Now consider the situation in which the hierarchy problem is solved by two or more large
extra dimensions of radius much larger than the inverse weak scale.  The cutoff of the
theory $M_*$ must then be at or about the weak scale if the Higgs mass is to be natural.
In our model above if the scalar field is localized to the same four brane in the higher
dimensional space as the Standard Model fields then $\lambda_1 \lambda_2 \le M_*^5$ is
automatically satisfied. However the form of the radion kinetic term is more complicated
because it can mix with other light scalar fields in higher dimensional gravity theory.
If there are N large extra dimensions whose radii $R_I(x)$ where $I$ runs from 1 to N
then there are mixing terms of the form $\partial_{\mu} R_I \partial^{\nu} r$ in the four
dimensional effective theory. However the coefficient of this term before Weyl rescaling
is of order $\frac{M_4^2}{rR_I}$ which is much less than the coefficient of the
$\partial_{\mu}r\partial^{\mu}r$ term which is of order $\frac{M_4^2}{r^2}$. Hence the
mixing terms are negligible and our conclusions remain unaffected.  But what if the
scalar field lives in the bulk? Then however a calculation analogous to the one above
shows that the radion potential is now of order $V_N M_*^{D+N} V$ where $V_N$ is the
volume of the large extra dimensions rather than just $M_*^D V$. This leads to a radion
mass which is much larger than inverse millimeter. Thus it is possible for the radion to
naturally be much heavier than inverse millimeter if it is stabilized by dynamics
involving the bulk of the large extra dimensions.

\section{Stabilization by Casimir Energy}

Finally for concreteness we consider the implications for short distance gravity
experiments of a model with one extra dimension stabilized using the Casimir energy of
the Standard Model fields which live in the bulk. Since this effect by itself is not
quite sufficient for stabilization we add a brane cosmological constant and bulk
cosmological constant{\footnote{Here we are ignoring the effects of brane localized
kinetic terms which we expect to be small.}}.

The contribution to the radion potential from the Casimir effect has been computed
{\cite{PP}}
\begin{equation}
\Omega_C\left(V\right) \approx -\frac{3N}{4\pi^2} \frac{1}{V^4}
\end{equation}

where the number $N$ depends on the number of fields in the bulk and the boundary
conditions imposed on them. The contribution to $N$ from a bulk vector field
is $3/2$ while for a bulk fermion $N = -2$. Every
real scalar contributes $1/2$ to $N$, while the graviton contributes $5/2$.
Thus
for the bulk Standard Model the total $N \approx -68 $.
The complete potential including the contributions from the bulk cosmological
constant and the brane localized cosmological constant has the form
\begin{equation}
\Omega\left(V\right) = \Omega_C\left(V\right) + \alpha V + \beta
\end{equation}
Minimizing this potential and requiring that the radion mass be heavier than 50
microns {\cite{A}} implies that the radius of the extra dimension must be smaller than
about
5 TeV$^{-1}$. While this result is sensitive not just to the Casimir energy term
but also the forms of additional terms in the radion potential, it shows that in
general scales as large as several TeV may be accesible to
sub-millimeter gravity experiments if the leading
contributions to the potential arise from Casimir energy.

\section{Conclusions}

We have investigated theories in which the Standard Model gauge fields propagate in extra
dimensions at the weak scale and shown that in a large class of models the radion remains
light and has a mass in the $10^{-3}$ eV range. The light radion is characteristic of
models in which stabilization does not involve scales larger than the cutoff of the
higher dimensional Standard Model, and in which the effective theory below the
compactification scale is four dimensional. This is generally the case when the hierarchy
problem is solved using an additional warped extra dimension and this class of models
will typically have a light radion unless the setup is fine tuned. However this is not
the case when the hierarchy problem is solved using large extra dimensions, and in such a
setup typically a light radion emerges only if the stabilization mechanism does not
involve the bulk of the large extra dimensions.

The light radion mediates an attractive force between matter that is comparable to but
somewhat stronger than gravity, and violates the Equivalence Principle albeit by a very
small amount. The strength of this force is a characteristic of the number of extra
dimensions and is less sensitive to other parameters such as the size and cutoff of the
extra dimensions. Current short diatance gravity experiments can probe extra dimensions
too small to be visible in collider experiments. In particular compactification radii as
small as 5 inverse TeV are accessible to current experiments if the extra dimensions are
stabilized using the Casimir energy of bulk Standard Model fields. We conclude that a
unique opportunity exists to explore physics at the weak scale by performing
gravitational experiments at the submillimeter scale.

\medskip

\noindent {\bf Acknowledgements} \\ We would like to thank Eduardo Ponton and Eric
Poppitz for discussions at various stages of this work.  The authors are supported by the
Director, Office of Science, Office of High Energy and Nuclear Physics, of the U. S.
Department of Energy under Contract DE-AC03-76SF00098, and by the National Science
Foundation under grant PHY-00-98840.

\end{document}